\documentclass[a4paper]{article}
\usepackage{natbib}
\usepackage{tgpagella}
\usepackage{geometry}
\usepackage{graphicx}

\title{Sensing and Modeling Human Behavior \\Using Social Media and Mobile Data}
\author{Abhinav Mehrotra and Mirco Musolesi \\ University College London and The Alan Turing Institute}
\date{}

\begin{document}
\maketitle

\begin{abstract}

In the past years we have witnessed the emergence of the new discipline of computational social science, which promotes a  new data-driven and computation-based approach to social sciences. In this article we discuss how the availability of new technologies such as online social media and mobile smartphones has allowed researchers to passively collect human behavioral data at a scale and a level of granularity that were just unthinkable some years ago. We also discuss how these digital traces can then be used to prove (or disprove) existing theories and develop new models of human behavior.
\end{abstract}

\bigskip
\noindent {\bf Keywords}: human behavior, mobile sensing, social media, computational social science, social physics, digital traces, mobility models, anticipatory mobile computing, mental health monitoring, smartphones.
\bigskip

\section{Introduction}
In the recent years, the emergence and widespread adoption of new technologies from social media to smartphones are rapidly changing the social sciences, since they have allowed researchers to analyze, study and model human behavior at a scale and at a granularity that were unthinkable just a few years ago. In fact, for example, social media allow us to collect information about ideas and opinions at very fine-grained temporal granularity and at a global scale~\citep{bond2012,leetaru2013mapping}; thanks to natural language processing~\citep{manning1999foundations}, we can also extract information about the emotional states of the individuals and model how people's opinion influence for example elections and other political events~\citep{williams2007social,gonzalez2013broadcasters}. Moreover, the advent of smartphones allow researchers and practitioners to collect data about human behavior using the sensors embedded in these devices, in particular GPS, accelerometers, microphones and cameras~\citep{internetcomputing}. By using this information, it is possible to model human behavior over space and time, trying to answer questions and prove (or disprove) new and traditional theories and models. For these reasons, researchers increasingly describe these developments as the emergence of \textit{computational social science}~\citep{lazer2009life}, i.e., a data-driven and computation-based way of performing social science research.

\begin{figure}[t]
\centering
\includegraphics[width=150mm]{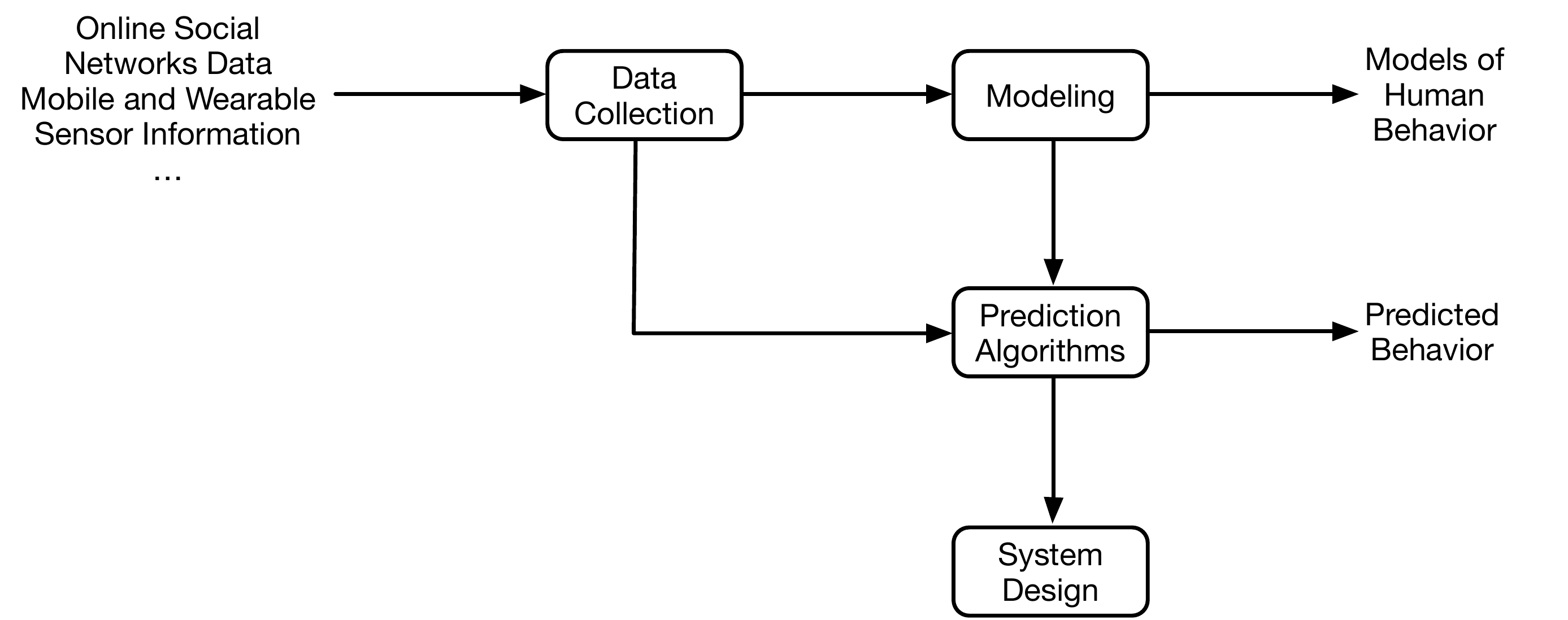}
\caption{Conceptual diagram of the steps involved in developing models and systems based on the analysis of human behavior.}
\label{fig1}
\end{figure}

The conceptual steps involved in modeling, analyzing and predicting human behavior are presented in Figure~\ref{fig1}. 
Firstly, we collect information about users' behavior from a variety of sources, e.g., social media and mobile sensor data. In the second step, the collected information is used to devise models and building prediction algorithms. It is also worth noting that models and prediction algorithms are also at the basis of the design of intelligent social systems, such as, for example, mobile applications, ubiquitous computing and Internet of Things systems and so on.   

In this article, we will discuss in particular two areas of this emerging discipline corresponding to two different types of data sources: the collection, analysis and modeling of data from online social media and mobile data. We will also provide some case studies by discussing some examples from recent projects conducted in this area. We will also discuss the impact that the availability of these technologies is having on the social sciences.

\section{Social Media}

Social media have been used to study a variety of research questions and problems, for example for understanding opinion formation~\citep{watts2007influentials,aral2012identifying}. Other interesting problems include the identification of influential spreaders of information~\citep{kitsak2010identification}, the maximization of the spreading process itself~\citep{kempe2003maximizing}, the impact of homophily in the diffusion of information~\citep{aral2009distinguishing} or the geographic distribution of human activities~\citep{noulas2011empirical}. Thanks to the availability of geo-social networks, it is also possible to identify the key spreaders in the geographic space as well~\citep{lima2012spatial}: this allows us to understand how specific people in a spatial network contribute to the dissemination of information and ideas in certain geographic areas. An interesting application is the diffusion of ideas and information in the social media through Web blogs and social media outlets such Facebook~\citep{ellison2007benefits} and Twitter~\citep{kwak2010twitter,cha2010measuring}. One of the first study of information diffusion in social media can be found in~\citep{gruhl2004information,adar2005tracking, leetaru2013mapping}. In~\citep{srep02980} the authors present a study regarding the modeling of the spreading of rumors related to the discovery of the Higgs boson~\citep{Aad20121,Chatrchyan201230} using Twitter data.  More in general, we have witnessed a large number of studies related to the analysis of topics discussed in Twitter. For example, in~\citep{romero2011differences} the authors analyze the adoption of hashtags, showing that politically controversial ones are particularly persistent.

The diffusion of online social network platforms has also led researchers to formulate hypotheses regarding the impact of connectivity and distance in a networked world~\citep{mok2010does}. In~\citep{scellato2010distance}, by analyzing four social networks, the authors show that distance is still a fundamental element in the establishment of social links, i.e., we should not probably talk about the \textit{death of distance}~\citep{cairncross2001death}, at least yet.  Hristova et al. showed that geo-social networks in different platforms are also highly correlated~\citep{hristova16:multilayer} and this fact can be used for effective link prediction, i.e., the establishing of a new connection in the social network.

Social media have also been used for the study of political events, such as for elections~\citep{williams2007social,tumasjan2010predicting, williams2012social,bond2012} and protests~\citep{gonzalez2011dynamics, gonzalez2013broadcasters, margetts2015political}. Through the analysis of social media it is possible to model the dynamics of election campaigns and map the public opinion of cities, regions and states. Finally, these new forms of data can also be used to study demographics and culture. Social media can be used to analyze the geo-demographic distribution of surnames for example~\citep{longley2016geo} or they can be used to study the diversity and gentrification phenomena~\citep{hristova16:measuring}. Finally, in~\citep{Silva14:youare} the authors discuss how data from the Foursquare platform can be used to understand cultural characteristics of geographic areas by looking at the presence of specific food and drinks venues.

Social media can be used to model events in real-time as they happen. For example, the Twitter data stream has been used to model natural events, such as earthquakes~\citep{sakaki2010earthquake}, and social phenomena, such as the dynamics in cities~\citep{zhang2013hoodsquare,xia2014citybeat}.

An interesting challenge in this area is related to modeling. The area of social networks has been investigated in the social sciences for many decades~\citep{wasserman1994social,borgatti2009network}. The novelty of these new data sources resides in the scale and level of detail offered by these networks.
The emergence area of network science~\citep{newman2010networks} offers a statistical variety of models that can be applied to the study of these networks. The goal of recent studies has been the development of specific techniques for  temporal~\citep{tang2010small,holme2012temporal}, spatiotemporal~\citep{Williams160196} and multi-layer networks~\citep{Hristova14:keep,boccaletti2014structure}.

Applications of these techniques include the modelling of spreading of information~\citep{srep02980} and the identification of the key players in a network for information diffusion and influence~\citep{gosh_2010_predictinginfluential,lima2012spatial}.  Network science techniques for identifying users that play key roles in their social network has several application in marketing~\citep{trusov2009effects} (usually these individuals are called ``influencers"), but also for example in epidemiology, i.e., for reaching out individuals in the community that might be able to inform the members of their social network about the availability of possible actions to take to contain an ongoing epidemics~\citep{lima2015disease}. Other applications include national security applications for example to target individuals that are at the center of criminal or terrorist networks~\citep{everton2012disrupting,campana2012listening}. For example, by analyzing the communication of individuals through different channels (cellular communications, social media and so on), it might be possible to reconstruct the network of these organization and the flow of information inside it. Indeed, these techniques are based on the analysis of the structure of the underlying networks and exploit graph theory results for identifying nodes that are at the center of potential communication paths between network nodes.

\section{Mobile Sensing} 

The mobile phone revolution has marked the beginning of the twenty-first century. Today mobile phones are the most ubiquitous personal computing devices in the planet covering around 96.8\% of the world population~\citep{ITUstats2013}. These devices have graduated over a period of time from merely calling instruments to \textit{smart and highly personal devices}. Besides being technical advanced and pervasive, these devices have a plethora of embedded sensing capabilities~\citep{internetcomputing}. For this reason, mobile phones have been increasingly used as an essential tool for the study of human behavior~\citep{pentland2014social}.

Sophisticated sensors embedded on a mobile phone are capable of generating high resolution data spanned over various context modalities capturing different aspects of human behavior, such as location, physical activity, audio environment, proximity with other objects, collocation with other devices, environmental indicators and many others~\citep{Mehrotra2014SenSocial}. At the same time, due to its pervasiveness and ability to passively log users' context, smartphones offer the opportunity of collecting data at an unprecedented scale. In other words, studies in numerous fields that were previously conducted at small scales through surveys can now be done at a large-scale and \textit{continuously} over time.

This possibility of collecting digital traces describing human behavior has revolutionized numerous fields. For example, access to human mobility data in real-time and at a large-scale provides enormous opportunities to improve environmental and traffic management~\cite{schneider2013unravelling,alexander2015origin} and understand the dynamics of cities~\citep{batty2013new}.
In computing the availability of the datasets has allowed researchers to monitor users' behavior~\citep{Lane2011BeWell,Canzian2015MoodTraces} or to inform the design of intelligent systems~\citep{Mehrotra2016PrefMiner}.

\subsection{Inferring Behavioral Information}
Scientists have explored the use of longitudinal context data for mining and uncovering users' everyday physical and cognitive behavior in natural settings. Most of the physical behavioral aspects, for example human mobility, can be directly inferred by mobile sensors. Scientists have demonstrated the use of data mining techniques to extract statistical information about users' physical fitness from mobile sensor data. Fitness tracking applications such as BeWell~\citep{Lane2011BeWell} and UbiFit~\citep{Consolvo2008} have shown the potential of mobile sensing for passively tracking different aspects of users' health, including physical activity, sleep duration and social isolation. These applications rely on sensors such as accelerometer, microphone and GPS, which are embedded in mobile phones, and they do not require any user input. This demonstrates the potential of mobile devices to empower users with the ability to self-monitor and curb certain physical fitness issues on their own. 

Moreover, studies have shown that cognitive context and health related information can be inferred as a function of physical context modalities. A variety of machine learning techniques can be used to analyze raw sensor data for modeling users' cognitive behavior. Unlike modeling physical behavior, cognitive context modeling requires the collection of labels of cognitive states (in the form of user feedback~\citep{Mehrotra2015ESM}) that are then mapped to raw sensor data and provided as inputs to machine learning tools for learning users' behavior. Therefore, passive sensing is not sufficient in these situations to capture the complexity of user's context.

EmotionSense~\citep{rachuri2010emotionsense} is an application that automatically recognizes the mood states of users by using voice classifiers running locally on phones. Similarly, the authors of~\citep{lu2012stresssense} propose a mechanism for unobtrusively recognizing the occurrence of stress by analyzing human voice captured through mobile phones. In~\citep{likamwa2013moodscope} the authors developed an application that infers users' mood based on mobile phone usage patterns; they suggest that the analysis of communication logs and application usage patterns can statistically infer users' daily mood average with a high accuracy.

Studies have also explored the potential of mobile sensing for monitoring mental health conditions~\citep{miller2012smartphone,bardram2013designing, lathia2013smartphones}. For example, the authors of~\citep{lacour2009tell} investigated the relationship between the changes in application usage behavior of patients and the onset of bipolar episodes. They show that the patterns of application usage can be correlated with different aspects of users' mood, sleep and irritability. The authors of another study~\citep{schleusing2011monitoring} have shown a strong correlation between physical activity and bipolar disorder, which can be used for an early intervention. In~\citep{Canzian2015MoodTraces} the authors show that the depressive states of users can be inferred purely from the mobility data collected via mobile phones. The authors suggest that characteristics of users' mobility are significantly correlated with their depressive moods. They show that machine learning models can successfully be used to predict changes in depressive states of users by using location data. 
It is worth noting that existing papers focus on correlation relationships, but the problem of causality is currently actively investigated by the research community~\citep{TsaMus15_investigating}.
The availability of these types of data sources will allow researchers to develop models that are able to capture not only behavioral features but also cognitive ones.

\begin{figure}[t!]
\centering
\includegraphics[width=90mm]{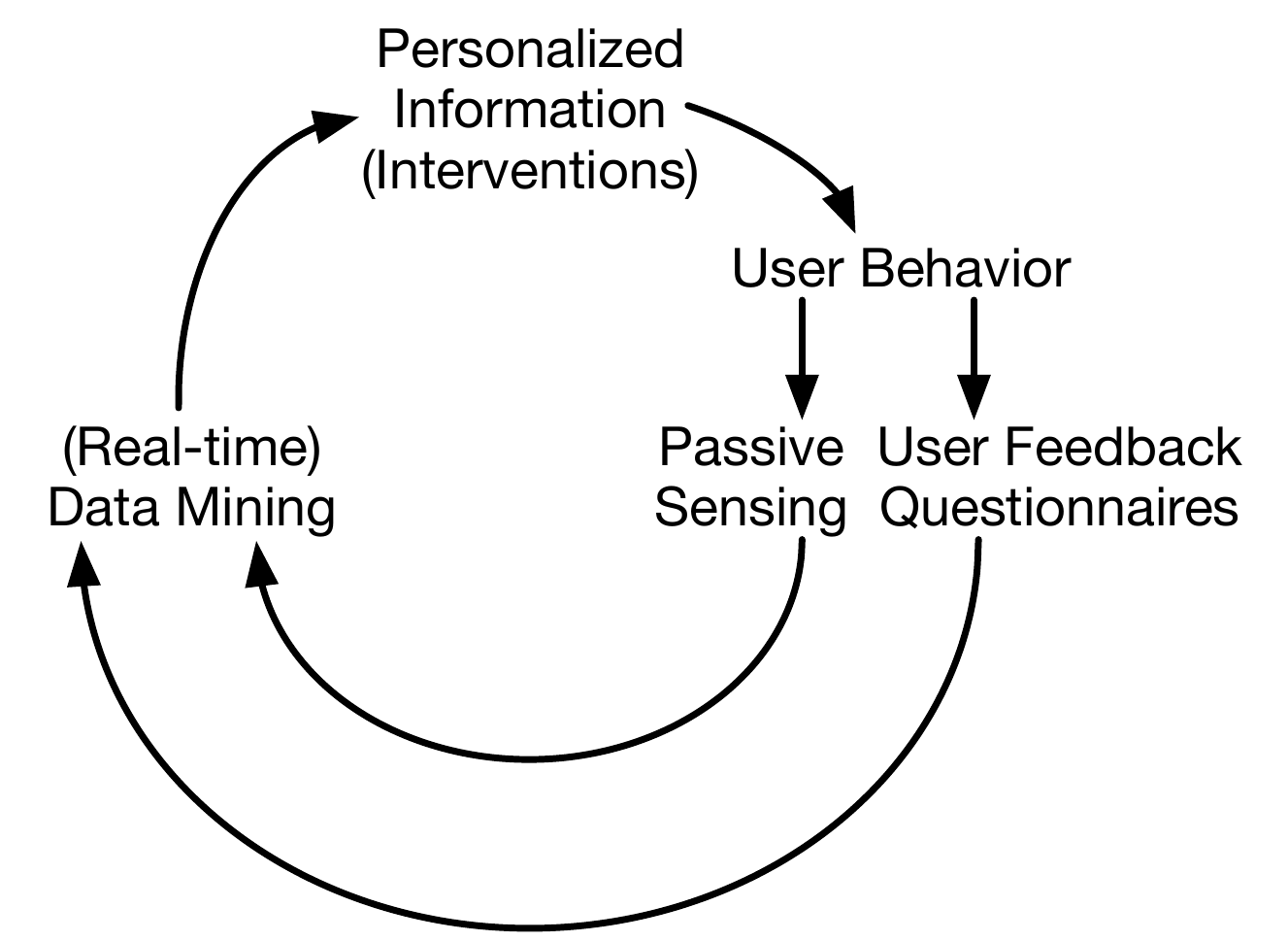}
\caption{General architectural model for systems based on the analysis and prediction of human behavior.}
\label{fig:mining}
\end{figure}

Furthermore, researchers have demonstrated the potential of feeding passively collected mobile sensing data into machine learning algorithms to build anticipatory applications~\citep{pejovic2015anticipatory}. An anticipatory application refers to an application that can act autonomously on the basis of past, present and predicted future context, and modify the future settings to satisfy users' requirements. Studies have demonstrated the potential of mobile sensing combined with predictive modeling techniques to anticipate contextual aspects such as mobile users' network connectivity~\citep{Choudhary2013PredictingStayAtHotspots}, communication patterns~\citep{Pielot2014Callavailability}, and application usage~\citep{Yan2012, Shin2012, MobileMiner2014}. In the recent years, researchers have proposed the use of sensed contextual information to build intelligent information delivery mechanism that learns users' preferences for receiving information in different situations~\citep{Mehrotra2015NotifyMe, Mehrotra2016MyPhoneAndMe, Mehrotra2016PrefMiner}. Figure~\ref{fig:mining} illustrates the high-level feedback loop that is at the basis of this class of systems.

For example, in~\citep{Mehrotra2016PrefMiner} Mehrotra et al. proposed a mechanism that learns user's preferences for receiving specific types of information on mobile phones in different contexts. 
Indeed, the design of the system reflects the steps presented in Figure~\ref{fig:mining}: sensing data, mining preferences of the individual, delivering personalized information based on his/her preferences. More specifically, firstly, the system continuously captures the individual's contextual data through passive mobile sensing and stores it in the memory. This information is periodically fed to a predictive algorithm. In some systems this step is executed in real-time, but in others it is executed periodically. In~\citep{Mehrotra2016PrefMiner} the data analysis is performed every night given the energy costs of the computation associated to it. A model of the user's preferences is derived from this analysis and it can be used to deliver targeted information of different types, such as personalized advertisement or messages for positive behavior change. The effect of the information dispatched to the users is then monitored through the sensors embedded in the devices. An example is the monitoring of the effects of a behavior intervention for weight management. A system might be able to check if the delivered information was effective by monitoring the activity level of a user through the accelerometer sensor embedded in the phone. Usually, in this type of systems users might also provide feedback (when asked) through questionnaire about the effectiveness of the system. For example, this is the case of the system presented in~\citep{Mehrotra2016PrefMiner}. This information is then used to adapt the behavior of the system itself, for example by refining the learning algorithms at the basis of it.

\subsection{Understanding Human Behavior}
One of the most informative sensing capability of mobile phones is the possibility of inferring users' location through Global Positioning System (GPS) or via Wi-Fi and GSM signals. 
The analysis of users' location data has led to numerous applications such as understanding the urban space~\citep{reades2007cellular}, improving its transportation network~\citep{alexander2015origin, ccolak2015analyzing, schneider2013unravelling}. Gonzalez et al.~\citep{gonzalez2008understanding} studied the trajectories of 100,000 anonymized users whose mobility was monitored via mobile phones for a time period of six months. Their study demonstrates that humans follow simple reproducible patterns with a high degree of temporal and spatial regularity. 

Another interesting aspect of human behavior is their social interaction, which can be captured for example through the combination of Bluetooth, location and communication logs. Ability to capture this information has made it possible for social scientists to conduct several ground-breaking studies to understand users' social interactions~\citep{choudhury2003sensing, eagle2009inferring,eagle2009eigenbehaviors,pentland2010honest,stehle2011high, madan2012sensing}. 
In~\citep{choudhury2003sensing} the authors demonstrate the potential of mobile wearable sensors to automatically and unobtrusively learn the social network structures that arise within human groups. They use these sensors to reliably estimate when people are close to others and when they are having a dialogue. Their study shows that user's centrality scores (i.e., a measure of influence and embeddedness of a user in the community) can also be computed by using raw sensor data.  In another pioneering study~\citep{stehle2011high} the authors suggest that characteristics of contact patterns within school children can be used as indicators for the propagation of diseases. They present a temporal evolution in the contact network of children in the school, which can be used to identify specific situations where children are in contact and during which infections may be transmitted. In another study~\citep{madan2012sensing} the authors exploit location and communication sensors to model sudden changes in the (long-term) health of individuals, as well as the dissemination of opinions in a community.

\section{Outlook}
			
In this article, we have discussed how the advent of new technologies, such as social media and smartphones, are transforming the way research in the social sciences is conducted. Indeed, by using these new technologies, it is possible to collect information about users at a level of granularity both in space and time that was just unthinkable some years ago. Researchers refer to these developments as the emergence of computational social science, i.e., a new data-driven and computation-based way of carrying out social science research.

In our opinion, we are just at the beginning of a revolution. As discussed in this article, social media and mobile data have already been used in several projects allowing researchers to address research questions in completely new ways. Many challenges have still to be tackled, especially from the methodological point of view. 
One of the most interesting ones is the possibility of integrating quantitative and qualitative methods starting from the availability of these new forms of data. Indeed, we believe that traditional methods in the social sciences should not just be replaced, but integrated in the process of development of new models and theories. In fact, for instance, data are usually collected in a passive way and, therefore, they are often hard to interpret. Classic qualitative methods might be used to understand the data, especially with respect to the limitations of the sources taken into consideration.

Moreover, another very important aspect of the problem is the bias associated with these data sources. We believe that the problems related to bias and sampling are the most significant in this area. In fact, the risk is to derive models that provide a non-realistic description of the reality or they fit the characteristics of very specific scenarios~\citep{lazer2014parable}. A typical example is the derivation of models from social media. The demographics of social media users do not usually reflect that of the entire population~\citep{duggan2013demographics}. In fact, for example, social media users are generally more educated and wealthy than the average population. Understand this bias is one of the most interesting questions in this research area. 

Finally, it is worth noting that the possibility of modeling human behavior at a fine-grained scale is having positive impact to many disciplines such as epidemiology~\citep{funk2010modelling,lima2015disease}. We believe that methods from computational social sciences will be more and more adopted in other areas. Most of the analysis performed by researchers is carried out using purpose-built software. However, we expect an increased integration of these methodologies in tools used by researchers such as traditional Geographic Information Systems~\citep{longley2015geographic}.

\section{Acknowledgements}

This work was supported by The Alan Turing Institute under the EPSRC grant EP/N510129/1 and through the EPSRC Grants ``MACACO: Mobile context-Adaptive CAching for Content-Centric networking'' (EP/L018829/2) and ``UPRISE-IoT: User-centric PRIvacy \& Security in IoT'' (EP/P016278/1).

\bibliographystyle{agsm}
\bibliography{biblio,bibliomirco}

\end{document}